\documentclass[notitlepage,a4paper,aps,prd,onecolumn,superscriptaddress,nofootinbib,groupedaddress,longbibliography]{revtex4-1}
\usepackage{amsmath}
\usepackage{amsfonts}
\usepackage{amssymb}
\usepackage[utf8]{inputenc}
\usepackage[T1]{fontenc}
\usepackage[colorlinks=true]{hyperref}
\usepackage{graphicx}
\usepackage{xcolor}
\usepackage[normalem]{ulem}

\newcommand{\diff}{\mathrm{d}}

\begin{document}

\title{Propagation of gravitational waves in symmetric teleparallel gravity theories}

\author{Manuel Hohmann}
\email{manuel.hohmann@ut.ee}
\affiliation{Laboratory of Theoretical Physics, Institute of Physics, University of Tartu, W. Ostwaldi 1, 50411 Tartu, Estonia}

\author{Christian Pfeifer}
\email{christian.pfeifer@ut.ee}
\affiliation{Laboratory of Theoretical Physics, Institute of Physics, University of Tartu, W. Ostwaldi 1, 50411 Tartu, Estonia}

\author{Jackson Levi Said}
\email{jsaid01@um.edu.mt}
\affiliation{Department of Physics, University of Malta, Msida, MSD 2080, Malta}

\author{Ulbossyn Ualikhanova}
\email{ulbossyn.ualikhanova@ut.ee}
\affiliation{Laboratory of Theoretical Physics, Institute of Physics, University of Tartu, W. Ostwaldi 1, 50411 Tartu, Estonia}

\begin{abstract}
Symmetric teleparallel gravity (STG) offers an interesting third geometric interpretation of gravitation besides its formulation in terms of a spacetime metric and Levi-Civita connection or its teleparallel formulation. It describes gravity through a connection which is not metric compatible, however is torsion and curvature free. We investigate the propagation velocity of the gravitational waves around Minkowski spacetime and their potential polarizations in a general class of STG theories, the so-called ``newer general relativity'' class. It is defined in terms of the most general Lagrangian that is quadratic in the nonmetricity tensor, does not contain its derivatives and is determined by five free parameters. In our work we employ the principal symbol method and the Newman-Penrose formalism, to find that all waves propagate with the speed of light, i.e., on the Minkowski spacetime light cone, and to classify the theories according to the number of polarizations of the waves depending on the choice of the parameters in the Lagrangian. In particular it turns out that there exist more theories than just the reformulation of general relativity which allow only for two polarization modes. We also present a visualization of the parameter space of the theory to better understand the structure of the model.
\end{abstract}

\maketitle

\section{Introduction}\label{sec:intro}
The observation of gravitational waves (GWs) has opened the possibility of a new window on strong field physics~\cite{Abbott:2016blz} that is not accessible by electromagnetic observations alone. While GW observations have continued to be confirmed, the first three-detector observation holds important significance in that such measurements allow for signal localization and, more to the purpose of this work, constraints on the six potential polarization modes of metric theories of gravity~\cite{Abbott:2017oio}. Moreover there has been the first multi messenger observations \cite{TheLIGOScientific:2017qsa} which constrain the difference of the propagation velocity between GW and electromagnetic waves in vacuum, which is different from zero in  various modified theories of gravity~\cite{Lombriser:2015sxa,Lombriser:2016yzn,Chakraborty:2017qve,Sakstein:2017xjx,Ezquiaga:2017ekz,Baker:2017hug,Akrami:2018yjz}. Thus GW observations offer the possibility for strong constraints on theories predicting extra modes and a propagation velocity different from the speed of light, and so may be the route to reducing the landscape of potential gravitational theories consistent with observation \cite{Clifton:2011jh}.

Viewed through the prism of the connection, metric theories of gravity can be classified into three broad classes of theories. The ones which use the Levi-Civita connection of the metric and its curvature, the ones which use the tetrads of a metric and their curvature free, metric-compatible, Weitzenb\"ock connection with torsion and the ones which use a curvature and torsion free symmetric teleparallel connection that is not metric compatible. This classification nicely highlights the sometimes overlooked point that curvature is a property of the connection and not of the metric tensor or the manifold \cite{Ortin:2015hya}. It becomes a property of the metric only through the use of the Levi-Civita connection. For the description of gravity it is remarkable that general relativity (GR) and the Einstein equations can be equivalently formulated in terms of either of the connections just mentioned \cite{Aldrovandi:2013wha,BeltranJimenez:2017tkd,Heisenberg:2018vsk}, i.e.\ all three connections can be used to define Lagrangians whose Euler-Lagrange equations coincide with the Einstein equations for a particular choice of contributing terms.

Historically most used for the construction of GR and extended theories of gravity \cite{Capozziello:2011et} is the Levi-Civita connection, resulting mainly in $f(R)$, $f(R,G)$ and similar theories. However, the use of torsion and nonmetricity allow for another kind of generalization \cite{BeltranJimenez:2018vdo}. In particular, the irreducible contributions of the Lagrangian of these two theories can be elevated to arbitrary coupling coefficients with a limit to their GR equivalent for a particular numerical choice. These two avenues of generalization are important because they may provide constraints on these novel and not extensively studied generalizations which may lead to a better understanding of the unique coincidence that GR appears to represent. Moreover, by altering the connection a new landscape of gravitational theories can be studied which differ from each other at a fundamental level in the classical regime \cite{Jarv:2018bgs}.

GWs offer the possibility of a model independent test of the polarization modes a theory exhibits \cite{Eardley:1973br,Eardley:1974nw}. In principle, this provides a strict test of which theories are realistic in the strong field regime. Thus far, the topic has not been studied as well for STG theories, as for torsion based (or teleparallel) gravity theories. In teleparallel gravity \cite{Cai:2015emx,Krssak:2015oua}, the propagation of GW modes has been shown to have a varied nature depending on the particular form the theory takes. This was first studied in Ref.\cite{Bamba:2013ooa} where it was found that the straightforward generalizations of the teleparallel equivalent of GR (TEGR), namely $f(T)$ theories, exhibit the same polarization structure as that of GR and thus is indistinguishable at the level of GW modes. The work then was further confirmed and expanded upon to encompass scalar fields and a generalized form of $f(R)$ gravity \cite{Abedi:2017jqx}, the speed of the GWs and the effect of the three-detector observation was then studied in Ref.\cite{Cai:2018rzd}, which then culminated in the explicit expression of the modes in these extended teleparallel theories in Ref.\cite{Farrugia:2018gyz}. In Ref.\cite{Hohmann:2018xnb,Hohmann:2018jso}, the general scenario of decomposed Lagrangians of both the torsional and nonmetricity situations is considered with clear groundwork for further analysis in either theory. Another approach to the propagator of generalized symmetric teleparallel gravity theories including higher derivative orders and making use of the Barnes-Rivers formalism can be found in Ref.\cite{Conroy:2017yln}.

In the present study, we investigate the GW polarization modes of the massless contribution in the general form of the STG setting. As in the teleparallel setting, since the Lagrangian can be divided into irreducible contributors, it is interesting to understand the GW mode structure that this seemingly arbitrary landscape provides \cite{BeltranJimenez:2017tkd,BeltranJimenez:2018vdo}. We then represent the resulting parameter space of this theory in a novel way, since the model has a lot of potential avenues to it.

The paper is organized as follows. In Sec.~\ref{sec:lintelep} we briefly introduce the key components of the model we are considering and form the linearized field equations. This is crucial to understanding the relevant contributions to the GW modes. In Fourier space, the field equations are then decomposed and the speed of GWs in STG is determine in Sec.~\ref{sec:speed} to determine the polarization states the Newman-Penrose formalism is considered in Sec.~\ref{sec:polar} where these states are depicted. Lastly, we close with a discussion in Sec.~\ref{sec:conclusion}.

\section{Linearized general symmetric teleparallel gravity theories}\label{sec:lintelep}
Before we derive the speed and polarization of gravitational waves in symmetric teleparallel gravity, we need to derive its linearized field equations. This is done in two parts. In section~\ref{ssec:geometry} we briefly review the underlying spacetime geometry and its gauge aspects. We turn our focus to the dynamics of the theory in section~\ref{ssec:action}, where we review the action and field equations, which we then linearize after gauge fixing.

\subsection{Geometry with nonmetricity}\label{ssec:geometry}
We start with a brief review of the underlying geometry involving nonmetricity, which we use in this article. The fundamental fields defining the geometry are a Lorentzian metric \(g_{\mu\nu}\) and an affine connection \(\Gamma^{\mu}{}_{\nu\rho}\). The connection is chosen to have vanishing curvature,
\begin{equation}\label{eqn:nocurv}
R^{\mu}{}_{\nu\rho\sigma} = \partial_{\rho}\Gamma^{\mu}{}_{\nu\sigma} - \partial_{\sigma}\Gamma^{\mu}{}_{\nu\rho} + \Gamma^{\mu}{}_{\omega\rho}\Gamma^{\omega}{}_{\nu\sigma} - \Gamma^{\mu}{}_{\omega\sigma}\Gamma^{\omega}{}_{\nu\rho} \equiv 0,
\end{equation}
and vanishing torsion
\begin{equation}\label{eqn:notors}
T^{\mu}{}_{\rho\sigma} = \Gamma^{\mu}{}_{\sigma\rho} - \Gamma^{\mu}{}_{\rho\sigma} \equiv 0\,.
\end{equation}
It does, however, possess in general non-vanishing nonmetricity,
\begin{equation}\label{eqn:nonmeten}
Q_{\alpha\mu\nu} = \nabla_{\alpha}g_{\mu\nu}\,.
\end{equation}
Indices are raised and lowered using the metric \(g_{\mu\nu}\). Note that due to the presence of nonmetricity this implies
\begin{equation}
Q_{\alpha}{}^{\mu\nu} = g^{\mu\rho}g^{\nu\sigma}Q_{\alpha\rho\sigma} = -\nabla_{\alpha}g^{\mu\nu}\,.
\end{equation}
The nonmetricity is obviously symmetric in its second and third index, \(Q_{\alpha\mu\nu} = Q_{\alpha\nu\mu}\), which allows the definition of two different traces,
\begin{equation}
Q_{\alpha} = g^{\mu\nu}Q_{\alpha\mu\nu}\,, \quad
\tilde{Q}_{\alpha} = g^{\mu\nu}Q_{\mu\nu\alpha}\,.
\end{equation}

The most general connection which satisfies the assumptions~\eqref{eqn:nocurv} and~\eqref{eqn:notors} is generated by a coordinate transformation defined by functions $\xi^{\mu}(x)$ in the form~\cite{Runkla:2018xrv,BeltranJimenez:2017tkd}
\begin{align}
\Gamma^\mu{}_{\nu\sigma} = \frac{\partial x^\mu}{\partial \xi^\rho}\partial_\nu \partial_\sigma \xi^\rho\,.
\end{align}
It further follows that it is always possible to find coordinates such that
\begin{equation}
\Gamma^\alpha_{\phantom{\alpha}\mu\nu} \equiv 0,
\end{equation}
not only at a single point, but in an open neighborhood. This particular choice of coordinates is known as the coincident gauge~\cite{BeltranJimenez:2018vdo}, and will be used throughout this work. Note that this uniquely determines the coordinate system \((x^{\mu})\) we use, up to linear transformations of the form
\begin{equation}
x^{\mu} \mapsto \tilde{\xi}^{\mu}(x)
= \tilde{\xi}^{\mu}(x_0) + (x^{\nu} - x_0^{\nu})\left.\partial_{\nu}\tilde{\xi}^{\mu}\right|_{x = x_0}\,,
\end{equation}
so that \(\partial_{\mu}\partial_{\nu}\tilde{\xi}^{\alpha} \equiv 0\). It follows that we have no further gauge freedom left to impose conditions on the metric degrees of freedom, except at a single point, as it is conventionally the case, e.g., in general relativity. In the coincident gauge covariant derivatives are replaced by partial derivatives, so that the nonmetricity reads
\begin{equation}
Q_{\alpha\mu\nu} = \partial_{\alpha}g_{\mu\nu}\,.
\end{equation}
We will make use of this formula in the following, when we derive the linearized field equations.


\subsection{Action and field equations}\label{ssec:action}
The starting point for the derivation of the linearized field equations is the ``newer general relativity'' action for the metric, the coordinate functions $\xi^\mu$ and the matter fields~\cite{Adak:2005cd,BeltranJimenez:2017tkd,BeltranJimenez:2018vdo}, which can be written in the form
\begin{equation}\label{eqn:action}
S[g_{\mu\nu}, \xi^\sigma, \chi^I] = S_g[g_{\mu\nu}, \xi^\sigma] + S_m[g_{\mu\nu}, \chi^I]\,, \quad
S_g = -\int_M\frac{\sqrt{-g}}{2}\mathbb{Q}\diff^4 x\,.
\end{equation}
We assume that the matter part \(S_m\) of the action does not depend on the affine connection \(\Gamma^{\alpha}{}_{\mu\nu}[\xi]\), but only on the metric \(g_{\mu\nu}\) and a set of matter fields \(\chi^I\). The gravitational part \(S_g\) of the action is expressed in terms of the nonmetricity scalar \(\mathbb{Q}\), seen as a function of the metric and the connection generating vector field, and is most conveniently defined via the nonmetricity conjugate
\begin{equation}\label{eqn:nmconjugate}
P^{\alpha}{}_{\mu\nu} = c_1Q^\alpha{}_{\mu\nu} + c_2Q_{(\mu}{}^{\alpha}{}_{\nu)} + c_3Q^{\alpha}g_{\mu\nu} + c_4\delta^{\alpha}_{(\mu}\tilde{Q}_{\nu)} + \frac{c_5}{2}\left(\tilde{Q}^{\alpha}g_{\mu\nu} + \delta^{\alpha}_{(\mu}Q_{\nu)}\right),
\end{equation}
as
\begin{equation}
\mathbb{Q} = Q_{\alpha}{}^{\mu\nu}P^{\alpha}{}_{\mu\nu}\,.
\end{equation}
This is the most general Lagrangian which is quadratic in the nonmetricity, unless one introduces also derivatives~\cite{Conroy:2017yln}. Choosing the parameters $c_1=-\frac{1}{4}, c_2=\frac{1}{2}, c_3=\frac{1}{4}, c_4=0$ and $c_5=-\frac{1}{2}$ one obtains the nonmetricity formulation of general relativity \cite{Nester:1998mp,BeltranJimenez:2017tkd}, which is usually called symmetric teleparallel equivalent of general relativity (STEGR).
By variation of the total action with respect to the metric, one obtains the field equations
\begin{equation}\label{eqn:feqmet}
\frac{2}{\sqrt{-g}}\nabla_{\alpha}(\sqrt{-g}P^{\alpha}{}_{\mu\nu}) + P_{\mu\sigma\rho}Q_{\nu}{}^{\sigma\rho} - 2Q_{\rho\mu}{}^{\sigma}P^{\rho}{}_{\nu\sigma} - \frac{1}{2}\mathbb{Q}g_{\mu\nu} = \mathfrak{T}_{\mu\nu}\,,
\end{equation}
where the energy-momentum tensor \(\mathfrak{T}_{\mu\nu}\) is derived from the matter action \(S_m\). To obtain the second set of field equations, we vary the total action with respect to the components of the connection generating coordinate functions $\xi^\mu$. Note that this is equivalent to performing a restricted variation of the flat, symmetric connection \(\Gamma^{\alpha}{}_{\mu\nu}\), which must be of the form \(\delta\Gamma^{\alpha}{}_{\mu\nu} = \nabla_{\mu}\nabla_{\nu}\delta\xi^{\alpha}\) in order to keep the vanishing torsion and curvature, \(\delta T^{\alpha}{}_{\mu\nu} \equiv 0\) and \(\delta R^{\alpha}{}_{\beta\mu\nu} \equiv 0\). After twice performing integration by parts, carefully taking into account the terms arising from \(\nabla_{\mu}\sqrt{-g}\) due to the nonmetricity, this yields the field equations
\begin{equation}\label{eqn:feqcon}
\nabla_{\mu}\nabla_{\nu}\left(\sqrt{-g}P^{\mu\nu}{}_{\alpha}\right) = 0\,.
\end{equation}
Note that their right hand side vanishes, since we have assumed no direct coupling of the matter to the flat, symmetric connection, and so the hypermomentum vanishes. We remark that this second set of field equations can alternatively be obtained from the diffeomorphism invariance of the gravitational action, giving an equivalent of the Bianchi identities, and the matter action, giving the matter energy-momentum conservation. This shows that the field equations~\eqref{eqn:feqmet} and~\eqref{eqn:feqcon} are not independent, and reflects the presence of the gauge symmetry under diffeomorphisms. Hence, we may restrict ourselves to solving the metric field equations~\eqref{eqn:feqmet}.

In order to linearize the metric field equations, we now adopt the coincident gauge \(\Gamma^{\alpha}{}_{\mu\nu} \equiv 0\) and consider a small perturbation around a Minkowski background metric,
\begin{equation}
g_{\mu\nu} = \eta_{\mu\nu} + h_{\mu\nu}\,.
\end{equation}
The nonmetricity tensor thus takes the form
\begin{equation}
Q_{\alpha\mu\nu} = \partial_{\alpha}h_{\mu\nu}\,.
\end{equation}
Further, we restrict ourselves to the vacuum field equations, so that \(\mathfrak{T}_{\mu\nu} \equiv 0\). Up to the linear order in the metric perturbations \(h_{\mu\nu}\), the metric field equations~\eqref{eqn:feqmet} then reduce to
\begin{equation}\label{eqn:linvaceom}
0 = 2c_1\square h_{\mu\nu} + (c_2 + c_4)\eta^{\alpha\sigma}\left(\partial_{\alpha}\partial_{\mu}h_{\sigma\nu} + \partial_{\alpha}\partial_{\nu}h_{\sigma\mu}\right) + 2c_3\eta_{\mu\nu}\eta^{\tau\omega}\square h_{\tau\omega} + c_5\eta_{\mu\nu}\eta^{\omega\gamma}\eta^{\alpha\sigma}\partial_{\alpha}\partial_{\omega}h_{\sigma\gamma} + c_5\eta^{\omega\sigma}\partial_{\mu}\partial_{\nu}h_{\omega\sigma}\,.
\end{equation}
Note that up to higher order terms, indices are now raised and lowered by the Minkowski metric \(\eta_{\mu\nu}\). This in particular applies to the d'Alembert operator $\Box = \eta^{\mu\nu}\partial_{\mu}\partial_{\nu}$. In the following, we will use the linearized equations~\eqref{eqn:linvaceom} in order to derive properties of gravitational wave propagation.

\section{Principal polynomial and speed of propagation}\label{sec:speed}
To determine the propagation speed of gravitational waves in nonmetricity theories of gravity we study the field equations in Fourier space. A necessary condition solutions of the field equations have to satisfy is, that the so called principal polynomial of the equations, as function of the wave covectors, has to vanish \cite{Hoermander1,Hoermander2}.

The field equations in Fourier space are
\begin{align}
	0 = \hat E_{\mu\nu} =
	\Big(
	2c_1\eta^{-1}(k,k) \delta^\lambda_\mu\delta^\rho_\nu + (c_2 + c_4)k^\lambda( k_\mu \delta^\rho_\nu + k_\nu \delta^\rho_\mu)
	+ 2 c_3 \eta_{\mu\nu}\eta^{-1}(k,k)\eta^{\lambda\rho} + c_5 (\eta_{\mu\nu}k^\lambda k^\rho + \eta^{\lambda\rho}k_\mu k_\nu )
	\Big)\hat h_{\lambda\rho}\,.
\end{align}
The principal polynomial of the equation is the determinant of the highest power in $k$ term. To calculate this determinant we decompose the equations with help of a decomposition with respect to a gauge vector field $\kappa^\mu$ which is dual to $k_\mu$, i.e.\ satisfies $\kappa^\mu k_\mu = 1$.

\begin{align}
	\hat h_{\lambda\rho} = S_{\lambda\rho} + 2 k_{(\lambda}V_{\rho)} + \frac{1}{3}(\eta_{\lambda\rho} - \tfrac{k_\lambda k_\rho}{\eta^{-1}(k,k)}) T + (k_\lambda k_\rho  -\tfrac{1}{4}\eta_{\lambda\rho}\eta^{-1}(k,k)) U\,,
\end{align}
where the divergence free symmetric traceless part $S_{\mu\nu}$ and the divergence free vector $V_\mu$ satisfy
\begin{align}
	\eta^{\lambda\rho}S_{\lambda\rho} = 0,\quad k^\lambda S_{\lambda\rho} = 0,\quad k^\rho V_\rho = 0\,.
\end{align}
The remaining scalars are the trace $T = \hat h_{\mu\nu}\eta^{\mu\nu}$ and the weighted double divergence $U = \frac{4}{3}\frac{h_{\mu\nu}k^\mu k^\nu}{\eta^{-1}(k,k)^{2}}$. Inserting this decomposition into the field equations yields
\begin{align}
0 &= \hat E_{\mu\nu}
=2c_1\eta^{-1}(k,k) S_{\mu\nu} + (2c_1 + c_2 + c_4)\eta^{-1}(k,k) 2 k_{(\mu}V_{\nu)} + \Big( \tfrac{2}{3}(c_1 + 3 c_3)\eta^{-1}(k,k)\eta_{\mu\nu} + (c_5 - \tfrac{2}{3}c_1) k_\mu k_\nu\Big)T \nonumber \\
&+ \Big( \tfrac{3}{4} ( c_5 - \tfrac{2}{3} c_1 )\eta^{-1}(k,k)^2\eta_{\mu\nu} + \tfrac{1}{2}(  4 c_1 + 3 c_2 + 3 c_4 ) \eta^{-1}(k,k) k_\mu k_\nu\Big)U
\end{align}
To further analyse them we consider their contractions with $k$, their trace and their symmetric traceless part
\begin{align}
0 &= \hat E_{\mu\nu} k^\mu k^\nu = (2 c_3 + c_5) \eta^{-1}(k,k)^2 T + (\tfrac{3}{4}c_5 + \tfrac{3}{2}(c_1 + c_2 + c_4))\eta^{-1}(k,k)^3 U\label{eq:ekk}\,,\\
0 &= \hat E^{\mu}{}_{\mu} =  (2 c_1 + 8 c_3 + c_5)\eta^{-1}(k,k)T + (3 c_5 + \tfrac{3}{2}(c_2 + c_4))\eta^{-1}(k,k)^2 U \label{eq:tr}\,,\\
0 &= \hat E_{\mu\nu} k^\mu - \tfrac{k_\nu}{\eta^{-1}(k,k)} E_{\rho\sigma}k^\rho k^\sigma =(2 c_1 + c_2 + c_4)\eta^{-1}(k,k)^2 V_\nu\,, \label{eq:vec}\\
0 &= \hat E_{\mu\nu} - \tfrac{1}{3} \big ( \eta_{\mu\nu} - \tfrac{k_\mu k_\mu}{\eta^{-1}(k,k)}\big)\hat E^{\sigma}{}_{\sigma} + \tfrac{1}{3}\big(\eta_{\mu\nu} - 4 \tfrac{k_\mu k_\nu}{\eta^{-1}(k,k)}\big)\tfrac{\hat E_{\rho\sigma}k^\rho k^\sigma}{\eta^{-1}(k,k)} =  2 c_1 \eta^{-1}(k,k) S_{\mu\nu}\,.\label{eq:symf}
\end{align}
To obtain the principal polynomial we can represent the decomposed equations as nearly diagonal matrix
\begin{align}\label{eq:matrixf}
\eta^{-1}(k,k)
\left(\begin{array}{cccc}
(2 c_3 + c_5)\eta^{-1}(k,k) & (\tfrac{3}{4}c_5 + \tfrac{3}{2}(c_1 + c_2 + c_4))\eta^{-1}(k,k)^2 & 0 & 0\\
(2 c_1 + 8 c_3 + c_5) & (3 c_5+\tfrac{3}{2}(c_2 + c_4))\eta^{-1}(k,k) & 0 & 0\\
0 & 0 & (2 c_1 + c_2 + c_4) \eta^{-1}(k,k) & 0\\
0 & 0 & 0 &  2 c_1 \\
\end{array}\right)
\left(\begin{array}{c}
T\\
U\\
V_\nu \\
S_{\mu\nu}\\
\end{array}\right)
=
\left(\begin{array}{c}
0\\
0\\
0 \\
0
\end{array}\right)\,,
\end{align}
and calculate its determinant
\begin{align}\label{eq:pp}
P(k) = (3 \times 2^3) c_1^5 (2 c_1 + c_2 + c_4)^3 (3 c_5^2 - 4 c_1^2 - 12 c_3 (c_2+c_4) - 4 c_1 (c_2 + c_4 + c_5 + 4 c_3) ) \eta^{-1}(k,k)^{15}\,.
\end{align}
The necessary and non-trivial condition, solutions of the field equations have to satisfy, is, that their wave covectors $k$ are such that $P(k) = 0$. From the above equation \eqref{eq:pp} we find that this implies $\eta^{-1}(k,k)=0$ must be satisfied, i.e.\ all propagating modes propagate on the null cone of the Minkowski metric, or in other words, with the vacuum speed of light. We like to remark that this does not mean that necessarily all the modes must be propagating degrees of freedom. The conclusion here is only that if they are, then they are propagating with the speed of light.

In case one considers nonmetricity gravity theories with parameters $c_1$ to $c_5$, such that one ore more field equations \eqref{eq:ekk} to \eqref{eq:symf} are solved trivially, for example $c_1 = 0$ for equation \eqref{eq:symf} or $2c_1 + c_2 + c_4 = 0$ for \eqref{eq:vec}, the corresponding modes, for example the tensor or vector mode, can not be propagating degrees of freedom of the theory. Their value must be defined by constraints which must be satisfied on initial data hypersurfaces. Such features become most visible in a full fledged Hamiltonian analysis of the theory in consideration, which shall be performed in the future.

To illustrate the statement just made we display the field equations for the values of the coefficients in the nonmetricity equivalent of general relativity $c_1=-\frac{1}{4}, c_2=\frac{1}{2}, c_3=\frac{1}{4}, c_4=0$ and $c_5=-\frac{1}{2}$
\begin{align}
0 &= 0\label{eq:ekkgr}\,,\\
0 &= \hat E^{\mu}{}_{\mu} =  \eta^{-1}(k,k)T - \tfrac{3}{4}\eta^{-1}(k,k)^2 U \label{eq:trgr}\,,\\
0 &= \hat E_{\mu\nu} k^\mu - \tfrac{k_\nu}{\eta^{-1}(k,k)} E_{\rho\sigma}k^\rho k^\sigma =0\,, \label{eq:vecgr}\\
0 &= \hat E_{\mu\nu} - \tfrac{1}{3} \big ( \eta_{\mu\nu} - \tfrac{k_\mu k_\mu}{\eta^{-1}(k,k)}\big)\hat E^{\sigma}{}_{\sigma} + \tfrac{1}{3}\big(\eta_{\mu\nu} - 4 \tfrac{k_\mu k_\nu}{\eta^{-1}(k,k)}\big)\tfrac{\hat E_{\rho\sigma}k^\rho k^\sigma}{\eta^{-1}(k,k)} =  \frac{1}{2} \eta^{-1}(k,k) S_{\mu\nu}\,.\label{eq:symfgr}
\end{align}
The vector modes $V_\mu$ can not be dynamical degrees of freedom since their field equation is satisfied identically. The two scalar modes are coupled and the tensor modes decouple. For general relativity it is know that a thorough Hamilton analysis yields that only two propagating degrees of remain and all other are fixed by constraints.

As final remark of this section we would like to remark here that, as in the analysis of linearized teleparallel theories of gravity \cite{Hohmann:2018xnb}, higher order poles appear in the propagators of the scalar and vector modes due do the higher then linear appearance of $\eta^{-1}(k,k)$ in the equations \eqref{eq:ekk}, \eqref{eq:tr} and \eqref{eq:vec}, which survive even in the non-metricity equivalent of general relativity for one of the scalar modes \eqref{eq:trgr}. On general grounds it is argued that the appearance of such terms signals the existence of ghost in the theory \cite{VanNieuwenhuizen:1973fi}. However the existence of such terms in the GR equivalent case shows that a more thorough analysis is required to identify if the ghost mode is coupling to the propagating field modes or not. The above mentioned complete Hamilton analysis of the theory considered here will also answer this question in the future.

\section{Newman-Penrose formalism and polarizations}\label{sec:polar}
We now focus on the polarization of gravitational waves. As we have seen in the previous section, gravitational waves in quadratic symmetric teleparallel gravity are described by Minkowski null waves, independently of the choice of the parameters \(c_1, \ldots, c_5\). This allows us to make use of the well-known Newman-Penrose formalism~\cite{Newman:1961qr} in order to decompose the linearized field equations into components, which directly correspond to particular polarizations. We then employ the classification scheme detailed in~\cite{Eardley:1973br,Eardley:1974nw}, which characterizes the allowed polarizations of gravitational waves in a given gravity theory by a representation of the little group, which is the two-dimensional Euclidean group \(\mathrm{E}(2)\) in case of null waves. In this section we determine the \(\mathrm{E}(2)\) class of quadratic symmetric teleparallel gravity for all possible values of the parameters \(c_1, \ldots, c_5\).

The main ingredient of the Newman-Penrose formalism is the choice of a particular complex double null basis of the tangent space. In the following, we will use the notation of~\cite{Will:1993ns} and denote the basis vectors by \(l^{\mu}, n^{\mu}, m^{\mu}, \bar{m}^{\mu}\). In terms of the canonical basis vectors of the Cartesian coordinate system they are defined as
\begin{equation}
l = \partial_0 + \partial_3\,, \quad
n = \frac{1}{2}(\partial_0 - \partial_3)\,, \quad
m = \frac{1}{\sqrt{2}}(\partial_1 + i\partial_2)\,, \quad
\bar{m} = \frac{1}{\sqrt{2}}(\partial_1 - i\partial_2)\,.
\end{equation}
We now consider a plane wave propagating in the positive \(x^3\) direction, which corresponds to a single Fourier mode. The wave covector then takes the form \(k_{\mu} = -\omega l_{\mu}\) and the metric perturbations can be written as
\begin{equation}\label{eqn:zwave}
h_{\mu\nu} = H_{\mu\nu}e^{i\omega u}\,,
\end{equation}
where we introduced the retarded time \(u = x^0 - x^3\) and the wave amplitudes are denoted \(H_{\mu\nu}\).

It follows from our choice of the matter coupling that test particles follow the geodesics of the metric, and hence the autoparallel curves of the Levi-Civita connection. The effect of a gravitational wave on an ensemble of test particles, or any other type of gravitational wave detector, therefore depends only on the Riemann tensor derived from the Levi-Civita connection. As shown in~\cite{Eardley:1974nw}, the Riemann tensor of a plane wave is determined completely by the six so-called electric components. For the wave~\eqref{eqn:zwave}, these can be written as
\begin{gather}
\Psi_2 = -\frac{1}{6}R_{nlnl} = \frac{1}{12}\ddot{h}_{ll}\,, \quad \Psi_3 = -\frac{1}{2}R_{nln\bar{m}} = -\frac{1}{2}\overline{R_{nlnm}} = \frac{1}{4}\ddot{h}_{l\bar{m}} = \frac{1}{4}\overline{\ddot{h}_{lm}}\,,\nonumber\\
\Psi_4 = -R_{n\bar{m}n\bar{m}} = -\overline{R_{nmnm}} = \frac{1}{2}\ddot{h}_{\bar{m}\bar{m}} = \frac{1}{2}\overline{\ddot{h}_{mm}}\,, \quad \Phi_{22} = -R_{nmn\bar{m}} = \frac{1}{2}\ddot{h}_{m\bar{m}}\,,\label{eqn:riemcomp}
\end{gather}
where dots denote derivatives with respect to \(u\). We now examine which of the components~\eqref{eqn:riemcomp} may occur for gravitational waves satisfying the linearized field equations~\eqref{eqn:linvaceom}.

\begin{figure}
\includegraphics[width=0.6\textwidth]{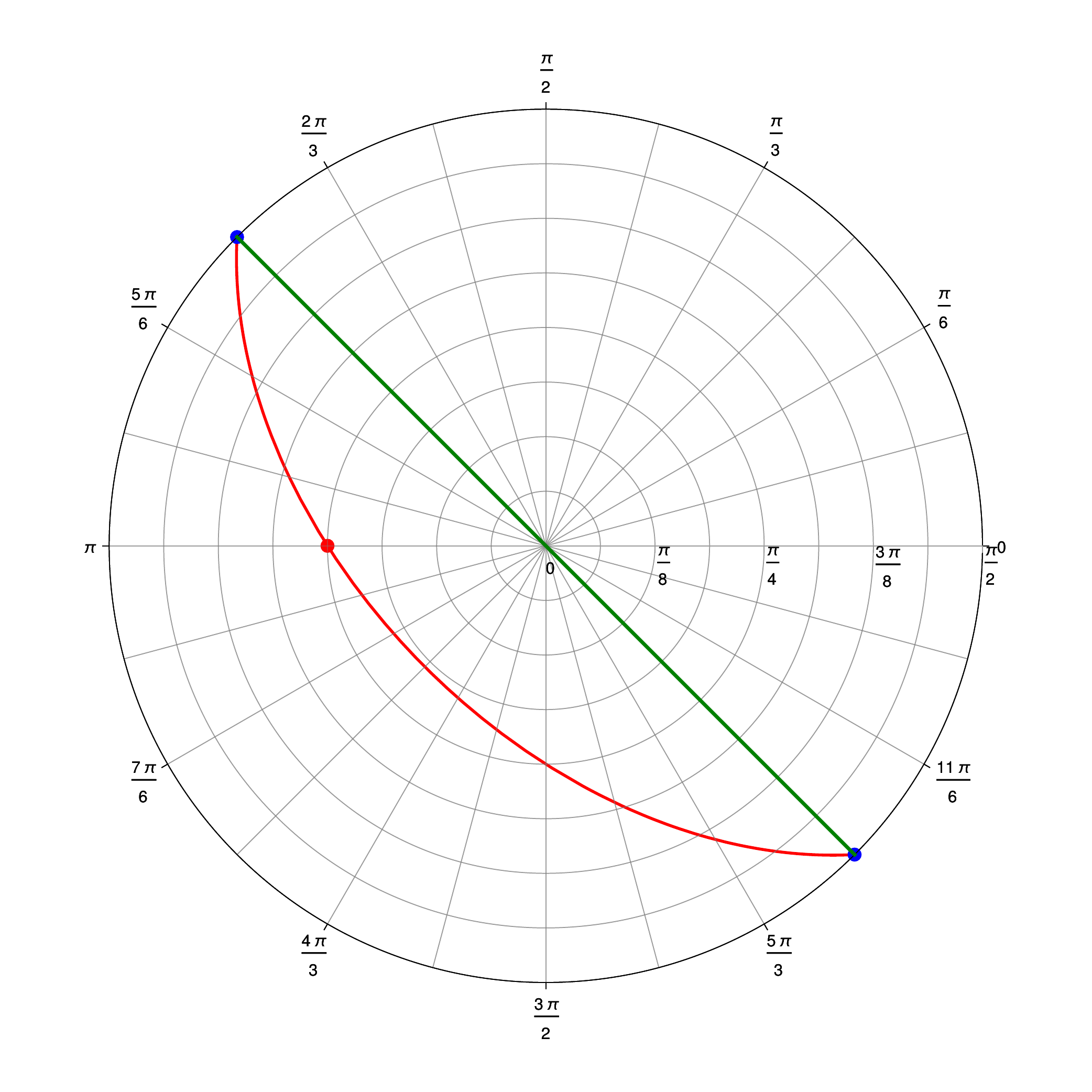}
\caption{(Color online.)
	 Visualization of the parameter space.
	 Blue Points: $c_2 + c_4 = c_5 = 0$, class \(\mathrm{II}_6\), 6 polarizations;
	 green line: $c_2 + c_4 = 0, c_5 \neq 0$, class \(\mathrm{III}_5\), 5 polarizations;
	 white area: $c_2 + c_4 \neq 0, c_2 + c_4 + c_5\neq 0$, class \(\mathrm{N}_3\), 3 polarizations;
	 red line: $c_2 + c_4 + c_5 = 0, c_5\neq 0$, class \(\mathrm{N}_2\), 2 polarizations.}
\label{fig:nmpol}
\end{figure}

Inserting the wave ansatz~\eqref{eqn:zwave} and writing the gravitational field strength tensor \(E_{\mu\nu}\) in the Newman-Penrose basis, we find that the five component equations
\begin{equation}
0 = E_{ll} = E_{lm} = E_{l\bar{m}} = E_{mm} = E_{\bar{m}\bar{m}}\,,
\end{equation}
are satisfied identically, while the remaining five component equations take the form
\begin{subequations}\label{eqn:np_wave}
\begin{align}
0 &= E_{nn} = 2c_5\ddot{h}_{m\bar{m}} - 2(c_2 + c_4 + c_5)\ddot{h}_{ln}\,,\label{eqn:np_nn}\\
0 &= E_{nm} = -(c_2 + c_4)\ddot{h}_{lm}\,,\label{eqn:np_nm}\\
0 &= E_{n\bar{m}} = -(c_2 + c_4)\ddot{h}_{l\bar{m}}\,,\label{eqn:np_nbm}\\
0 &= E_{m\bar{m}} = c_5\ddot{h}_{ll}\,,\label{eqn:np_mbm}\\
0 &= E_{ln} = -(c_2 + c_4)\ddot{h}_{ll}\,.\label{eqn:np_ln}
\end{align}
\end{subequations}
Note in particular that the parameters \(c_1\) and \(c_3\) do not appear in these equations. This can be understood by taking a closer look at the linearized field equations~\eqref{eqn:linvaceom}. Here the constants \(c_1\) and \(c_3\) appear in front of terms of the form \(\square h_{\mu\nu}\), where \(\square = \eta^{\mu\nu}\partial_{\mu}\partial_{\nu}\) is the d'Alembert operator of the flat background. These terms vanish identically for the null wave~\eqref{eqn:zwave}, independently of the amplitudes \(H_{\mu\nu}\), since the retarded time \(u\) is a light cone coordinate, and so \(\square e^{i\omega u} \equiv 0\). This can also be seen from the fact that the corresponding wave covector \(k_{\mu} = -\omega l_{\mu}\) is null, i.e., \(\eta^{-1}(k,k) = 0\), which is a necessary condition for solving the equations~\eqref{eqn:linvaceom} as shown in the preceding section~\ref{sec:speed}. Hence, it is a direct consequence of the form of the propagator that the allowed polarizations depend only on the remaining parameters \(c_2, c_4, c_5\). We now distinguish the following cases, which are also visualized in the diagram in figure~\ref{fig:nmpol} which we explain later in this section:
\begin{itemize}
\item
\(c_2 + c_4 = c_5 = 0\):
In this case equations~\eqref{eqn:np_mbm} and~\eqref{eqn:np_ln} are satisfied identically for arbitrary amplitudes \(H_{ll}\). For waves of this type the corresponding component \(R_{nlnl} = -6\Psi_2\) of the Riemann tensor, which describes a longitudinally polarized wave mode, is allowed to be nonzero. Following the classification detailed in~\cite{Eardley:1974nw}, they belong to the \(\mathrm{E}(2)\) class \(\mathrm{II}_6\) with six polarizations. This case corresponds to the two blue points in figure~\ref{fig:nmpol}, which is actually a line in the three-dimensional parameter space, and hence a single point in the projected parameter space shown in the diagram, which happens to lie on the cut \(c_5 = 0\) and hence appears twice on the circular perimeter.

\item
\(c_2 + c_4 = 0\) and \(c_5 \neq 0\):
It follows from the second condition that equation~\eqref{eqn:np_mbm} prohibits a non-vanishing amplitude \(H_{ll}\). Hence, there is no longitudinal mode \(\Psi_2\). Equations~\eqref{eqn:np_nm} and~\eqref{eqn:np_nbm} are satisfied identically for arbitrary amplitudes \(H_{lm}\) and \(H_{l\bar{m}}\). It then follows that \(R_{nln\bar{m}} = -2\Psi_3\), whose complex components describe two vector polarizations, is allowed to be nonzero. Waves of this type belong to the \(\mathrm{E}(2)\) class \(\mathrm{III}_5\), and there are five polarizations. This case is represented by the green line in figure~\ref{fig:nmpol}.

\item
\(c_2 + c_4 \neq 0\) and \(c_2 + c_4 + c_5 \neq 0\):
In this case it follows from equations~\eqref{eqn:np_nm}, \eqref{eqn:np_nbm} and~\eqref{eqn:np_ln} that \(H_{ll}\), \(H_{lm}\) and \(H_{l\bar{m}}\) must vanish. Hence, the longitudinal mode \(\Psi_2\) and vector modes \(\Psi_3\) are prohibited. The remaining linearized field equation which allows for non-vanishing solutions is equation~\eqref{eqn:np_nn}. In particular, it allows for a non-vanishing amplitude \(H_{m\bar{m}}\), and hence a non-vanishing component \(R_{nmn\bar{m}} = -\Phi_{22}\) of the Riemann tensor. The corresponding scalar wave mode is called the breathing mode. This wave has the \(\mathrm{E}(2)\) class \(\mathrm{N}_3\), exhibiting three polarizations. Almost all points of the parameter space, shown in white in figure~\ref{fig:nmpol}, belong to this class.

\item
\(c_2 + c_4 + c_5 = 0\) and \(c_5 \neq 0\):
The linearized field equations~\eqref{eqn:np_wave} in the Newman-Penrose basis now yield the conditions \(H_{ll} = H_{lm} = H_{l\bar{m}} = H_{m\bar{m}} = 0\). It thus follows that the longitudinal mode \(\Psi_2\), the vector modes \(\Psi_3\) and also the breathing mode \(\Phi_{22}\) must vanish. The only unrestricted electric components of the Riemann tensor are therefore \(R_{nmnm} = -\bar{\Psi}_4\) and its complex conjugate, corresponding to two tensor modes. The \(\mathrm{E}(2)\) class of this wave is \(\mathrm{N}_2\), so that there are two polarizations. This case is shown as a red line in figure~\ref{fig:nmpol}. Note in particular that STEGR, marked as a red point, belongs to this class, as one would expect.
\end{itemize}

We have visualized the aforementioned cases in figure~\ref{fig:nmpol}, which we constructed as follows. We first made the assumption that at least one of the parameters \(c_2, c_4, c_5\) is non-vanishing and introduced normalized parameters
\begin{equation}
\tilde{c}_i = \frac{c_i}{\sqrt{c_2^2 + c_4^2 + c_5^2}}
\end{equation}
for \(i = 2, 4, 5\). One easily checks that the \(\mathrm{E}(2)\) classes we found only depend on these normalized parameters, except for the case \(c_2 = c_4 = c_5 = 0\) belonging to class \(\mathrm{II}_6\). We then introduced polar coordinates \((\theta, \phi)\) on the unit sphere to express the parameters as
\begin{equation}
\tilde{c}_2 = \sin\theta\cos\phi\,, \quad
\tilde{c}_4 = \sin\theta\sin\phi\,, \quad
\tilde{c}_5 = \cos\theta\,.
\end{equation}
Since the \(\mathrm{E}(2)\) class is the same for antipodal points on the parameter sphere, we restrict ourselves to the hemisphere \(\tilde{c}_5 \geq 0\), and hence \(0 \leq \theta \leq \frac{\pi}{2}\); this is equivalent to identifying antipodal points on the sphere and working with the projective sphere instead, provided that we also identify antipodal points on the equator \(\tilde{c}_5 = 0\). We then considered \((\theta, \phi)\) as polar coordinates on the plane in order to draw the diagram shown in figure~\ref{fig:nmpol}. Note that antipodal points on the perimeter, such as the two blue points, are identified with each other, since they describe the same class of theories; in fact, these blue points correspond to a straight line passing through and including the origin \(c_2 = c_4 = c_5 = 0\).

This concludes our discussion of gravitational wave polarizations. We have seen that the parameters \(c_1, c_3\) have no influence on the allowed polarizations, while depending on the parameters \(c_2, c_4, c_5\) we obtain the \(\mathrm{E}_2\) class \(\mathrm{II}_6\), \(\mathrm{III}_5\), \(\mathrm{N}_3\) or \(\mathrm{N}_2\), with \(\mathrm{N}_3\) filling most of the parameter space. We have also seen that there exists a four parameter family of theories besides STEGR which is of class \(\mathrm{N}_2\) and thus exhibits the same two tensor modes as in general relativity. Theories in this class therefore cannot be distinguished from general relativity by observing the polarizations of gravitational waves alone.

\section{Conclusion}\label{sec:conclusion}
We studied the propagation of gravitational waves in the most general class of symmetric teleparallel gravity theories whose action is quadratic in the nonmetricity tensor. The wave we considered is modeled as a linear perturbation of a Minkowski background metric in the coincident gauge, in which the coefficients of the flat, symmetric connection vanish. We derived the principal polynomial of the linearized field equations and found that gravitational waves propagate at the speed of light, i.e., their wave covector must be given by a null covector of the Minkowski spacetime background. Further, we made use of the Newman-Penrose formalism to derive the possible polarizations of gravitational waves. Our results show that the two tensor polarizations, which are present also in general relativity, are allowed for the whole class of theories we considered, while additional modes - two vector modes and up to two scalar modes - may be present for particular models within this class. We found that the symmetric teleparallel equivalent of general relativity is not the unique theory exhibiting exactly two polarizations, but there is a four parameter family of theories with the same property. It thus follows that observations of gravitational wave polarizations may only give partial results on the parameter space of these theories.

We remark that although we restricted our analysis to theories whose action is quadratic in the nonmetricity tensor, our results are valid for a significantly larger class of theories. This is due to the fact that the nonmetricity is linear in the metric perturbations, so that the action is already quadratic in the perturbations. Hence, any higher order correction terms would have no influence on the linearized field equations. This is shown, e.g., in~\cite{Hohmann:2018xnb} for the polarizations of gravitational waves in a more general class of theories, whose Lagrangian is defined by a free function of the five scalar terms quadratic in nonmetricity considered in this article.

Another possible class of extensions is to consider additional fields non-minimally coupled to nonmetricity and to study their influence both on the speed and the polarization of gravitational waves. A canonical example is given by scalar-nonmetricity theories~\cite{Jarv:2018bgs,Runkla:2018xrv} constructed from the STEGR nonmetricity scalar and an additional scalar field, where one would expect the presence of an additional scalar mode compared to general relativity as it is also the case for scalar-curvature gravity. These theories can be extended further by replacing the STEGR nonmetricity scalar with the general quadratic nonmetricity scalar which defined the Lagrangian considered in this article.

Finally, another interesting extension of our work would be to study gravitational waves as a perturbation to a Friedmann-Lemaitre-Robertson-Walker metric. One may expect that in this case also nonmetricity terms of higher then quadratic order in the Lagrangian would affect the result, as they would lead to modifications of the background dynamics. This modified expansion history might thus also leave an imprint on the observed gravitational waves propagating in a cosmological background.

In conclusion, the formulation of theories of gravity in the symmetric teleparallel/nonmetricity language allows for promising extensions of GR which are consistent with the basic gravitational wave observations. An analysis of further observables in this particular class of theories, like the calculation of PPN parameters, rotational curves of galaxies and the cosmological expansion of the universe, will explore their viability further in the future.

\begin{acknowledgments}
The authors thank Lavinia Heisenberg for helpful comments, and Ott Vilson for helpful feedback and correcting an error in the original version. They were supported by the Estonian Ministry for Education and Science through the Institutional Research Support Project IUT02-27 and Startup Research Grant PUT790, as well as the European Regional Development Fund through the Center of Excellence TK133 ``The Dark Side of the Universe''. This article is based upon work from COST Action CANTATA, supported by COST (European Cooperation in Science and Technology).
\end{acknowledgments}

\bibliographystyle{utphys}
\bibliography{wavesNM}

\providecommand{\href}[2]{#2}\begingroup\raggedright\begin{thebibliography}{10}

\bibitem{Abbott:2016blz}
{\bf Virgo, LIGO Scientific} Collaboration, B.~P. Abbott {\em et al.},
  ``{Observation of Gravitational Waves from a Binary Black Hole Merger},''
  \href{http://dx.doi.org/10.1103/PhysRevLett.116.061102}{{\em Phys. Rev.
  Lett.} {\bf 116} (2016) no.~6, 061102},
\href{http://arxiv.org/abs/1602.03837}{{\tt arXiv:1602.03837 [gr-qc]}}.

\bibitem{Abbott:2017oio}
{\bf Virgo, LIGO Scientific} Collaboration, B.~P. Abbott {\em et al.},
  ``{GW170814: A Three-Detector Observation of Gravitational Waves from a
  Binary Black Hole Coalescence},''
  \href{http://dx.doi.org/10.1103/PhysRevLett.119.141101}{{\em Phys. Rev.
  Lett.} {\bf 119} (2017) no.~14, 141101},
\href{http://arxiv.org/abs/1709.09660}{{\tt arXiv:1709.09660 [gr-qc]}}.

\bibitem{TheLIGOScientific:2017qsa}
{\bf Virgo, LIGO Scientific} Collaboration, B.~Abbott {\em et al.},
  ``{GW170817: Observation of Gravitational Waves from a Binary Neutron Star
  Inspiral},'' \href{http://dx.doi.org/10.1103/PhysRevLett.119.161101}{{\em
  Phys. Rev. Lett.} {\bf 119} (2017) no.~16, 161101},
\href{http://arxiv.org/abs/1710.05832}{{\tt arXiv:1710.05832 [gr-qc]}}.

\bibitem{Lombriser:2015sxa}
L.~Lombriser and A.~Taylor, ``{Breaking a Dark Degeneracy with Gravitational
  Waves},'' \href{http://dx.doi.org/10.1088/1475-7516/2016/03/031}{{\em JCAP}
  {\bf 1603} (2016) no.~03, 031},
\href{http://arxiv.org/abs/1509.08458}{{\tt arXiv:1509.08458 [astro-ph.CO]}}.

\bibitem{Lombriser:2016yzn}
L.~Lombriser and N.~A. Lima, ``{Challenges to Self-Acceleration in Modified
  Gravity from Gravitational Waves and Large-Scale Structure},''
  \href{http://dx.doi.org/10.1016/j.physletb.2016.12.048}{{\em Phys. Lett.}
  {\bf B765} (2017)  382--385},
\href{http://arxiv.org/abs/1602.07670}{{\tt arXiv:1602.07670 [astro-ph.CO]}}.

\bibitem{Chakraborty:2017qve}
S.~Chakraborty, K.~Chakravarti, S.~Bose, and S.~SenGupta, ``{Signatures of
  extra dimensions in gravitational waves from black hole quasinormal modes},''
  \href{http://dx.doi.org/10.1103/PhysRevD.97.104053}{{\em Phys. Rev.} {\bf
  D97} (2018) no.~10, 104053},
\href{http://arxiv.org/abs/1710.05188}{{\tt arXiv:1710.05188 [gr-qc]}}.

\bibitem{Sakstein:2017xjx}
J.~Sakstein and B.~Jain, ``{Implications of the Neutron Star Merger GW170817
  for Cosmological Scalar-Tensor Theories},''
  \href{http://dx.doi.org/10.1103/PhysRevLett.119.251303}{{\em Phys. Rev.
  Lett.} {\bf 119} (2017) no.~25, 251303},
\href{http://arxiv.org/abs/1710.05893}{{\tt arXiv:1710.05893 [astro-ph.CO]}}.

\bibitem{Ezquiaga:2017ekz}
J.~M. Ezquiaga and M.~Zumalacárregui, ``{Dark Energy After GW170817: Dead Ends
  and the Road Ahead},''
  \href{http://dx.doi.org/10.1103/PhysRevLett.119.251304}{{\em Phys. Rev.
  Lett.} {\bf 119} (2017) no.~25, 251304},
\href{http://arxiv.org/abs/1710.05901}{{\tt arXiv:1710.05901 [astro-ph.CO]}}.

\bibitem{Baker:2017hug}
T.~Baker, E.~Bellini, P.~G. Ferreira, M.~Lagos, J.~Noller, and I.~Sawicki,
  ``{Strong constraints on cosmological gravity from GW170817 and GRB
  170817A},'' \href{http://dx.doi.org/10.1103/PhysRevLett.119.251301}{{\em
  Phys. Rev. Lett.} {\bf 119} (2017) no.~25, 251301},
\href{http://arxiv.org/abs/1710.06394}{{\tt arXiv:1710.06394 [astro-ph.CO]}}.

\bibitem{Akrami:2018yjz}
Y.~Akrami, P.~Brax, A.-C. Davis, and V.~Vardanyan, ``{Neutron star merger
  GW170817 strongly constrains doubly coupled bigravity},''
  \href{http://dx.doi.org/10.1103/PhysRevD.97.124010}{{\em Phys. Rev.} {\bf
  D97} (2018) no.~12, 124010},
\href{http://arxiv.org/abs/1803.09726}{{\tt arXiv:1803.09726 [astro-ph.CO]}}.

\bibitem{Clifton:2011jh}
T.~Clifton, P.~G. Ferreira, A.~Padilla, and C.~Skordis, ``{Modified Gravity and
  Cosmology},'' \href{http://dx.doi.org/10.1016/j.physrep.2012.01.001}{{\em
  Phys. Rept.} {\bf 513} (2012)  1--189},
\href{http://arxiv.org/abs/1106.2476}{{\tt arXiv:1106.2476 [astro-ph.CO]}}.

\bibitem{Ortin:2015hya}
T.~Ortin, \href{http://dx.doi.org/10.1017/CBO9781139019750}{{\em {Gravity and
  Strings}}}.
\newblock Cambridge Monographs on Mathematical Physics. Cambridge University
  Press, 2015.
\newblock
\url{http://www.cambridge.org/mw/academic/subjects/physics/theoretical-physics-and-mathematical-physics/gravity-and-strings-2nd-edition}.
\newblock

\bibitem{Aldrovandi:2013wha}
R.~Aldrovandi and J.~G. Pereira,
  \href{http://dx.doi.org/10.1007/978-94-007-5143-9}{{\em {Teleparallel
  Gravity}}}, vol.~173.
\newblock Springer, Dordrecht,
2013.
\newblock

\bibitem{BeltranJimenez:2017tkd}
J.~Beltrán~Jiménez, L.~Heisenberg, and T.~Koivisto, ``{Coincident General
  Relativity},'' \href{http://dx.doi.org/10.1103/PhysRevD.98.044048}{{\em Phys.
  Rev.} {\bf D98} (2018) no.~4, 044048},
\href{http://arxiv.org/abs/1710.03116}{{\tt arXiv:1710.03116 [gr-qc]}}.

\bibitem{Heisenberg:2018vsk}
L.~Heisenberg, ``{A systematic approach to generalisations of General
  Relativity and their cosmological implications},''
  \href{http://dx.doi.org/10.1016/j.physrep.2018.11.006}{{\em Phys. Rept.} {\bf
  796} (2019)  1--113},
\href{http://arxiv.org/abs/1807.01725}{{\tt arXiv:1807.01725 [gr-qc]}}.

\bibitem{Capozziello:2011et}
S.~Capozziello and M.~De~Laurentis, ``{Extended Theories of Gravity},''
  \href{http://dx.doi.org/10.1016/j.physrep.2011.09.003}{{\em Phys. Rept.} {\bf
  509} (2011)  167--321},
\href{http://arxiv.org/abs/1108.6266}{{\tt arXiv:1108.6266 [gr-qc]}}.

\bibitem{BeltranJimenez:2018vdo}
J.~Beltrán~Jiménez, L.~Heisenberg, and T.~S. Koivisto, ``{Teleparallel
  Palatini theories},''
  \href{http://dx.doi.org/10.1088/1475-7516/2018/08/039}{{\em JCAP} {\bf 1808}
  (2018)  039},
\href{http://arxiv.org/abs/1803.10185}{{\tt arXiv:1803.10185 [gr-qc]}}.

\bibitem{Jarv:2018bgs}
L.~Järv, M.~Rünkla, M.~Saal, and O.~Vilson, ``{Nonmetricity formulation of
  general relativity and its scalar-tensor extension},''
  \href{http://dx.doi.org/10.1103/PhysRevD.97.124025}{{\em Phys. Rev.} {\bf
  D97} (2018) no.~12, 124025},
\href{http://arxiv.org/abs/1802.00492}{{\tt arXiv:1802.00492 [gr-qc]}}.

\bibitem{Eardley:1973br}
D.~M. Eardley, D.~L. Lee, A.~P. Lightman, R.~V. Wagoner, and C.~M. Will,
  ``{Gravitational-wave observations as a tool for testing relativistic
  gravity},''
\href{http://dx.doi.org/10.1103/PhysRevLett.30.884}{{\em Phys. Rev. Lett.} {\bf
  30} (1973)  884--886}.

\bibitem{Eardley:1974nw}
D.~M. Eardley, D.~L. Lee, and A.~P. Lightman, ``{Gravitational-wave
  observations as a tool for testing relativistic gravity},''
\href{http://dx.doi.org/10.1103/PhysRevD.8.3308}{{\em Phys. Rev.} {\bf D8}
  (1973)  3308--3321}.

\bibitem{Cai:2015emx}
Y.-F. Cai, S.~Capozziello, M.~De~Laurentis, and E.~N. Saridakis, ``{f(T)
  teleparallel gravity and cosmology},''
  \href{http://dx.doi.org/10.1088/0034-4885/79/10/106901}{{\em Rept. Prog.
  Phys.} {\bf 79} (2016) no.~10, 106901},
\href{http://arxiv.org/abs/1511.07586}{{\tt arXiv:1511.07586 [gr-qc]}}.

\bibitem{Krssak:2015oua}
M.~Krššák and E.~N. Saridakis, ``{The covariant formulation of f(T)
  gravity},'' \href{http://dx.doi.org/10.1088/0264-9381/33/11/115009}{{\em
  Class. Quant. Grav.} {\bf 33} (2016) no.~11, 115009},
\href{http://arxiv.org/abs/1510.08432}{{\tt arXiv:1510.08432 [gr-qc]}}.

\bibitem{Bamba:2013ooa}
K.~Bamba, S.~Capozziello, M.~De~Laurentis, S.~Nojiri, and D.~Sáez-Gómez,
  ``{No further gravitational wave modes in $F(T)$ gravity},''
  \href{http://dx.doi.org/10.1016/j.physletb.2013.10.022}{{\em Phys. Lett.}
  {\bf B727} (2013)  194--198},
\href{http://arxiv.org/abs/1309.2698}{{\tt arXiv:1309.2698 [gr-qc]}}.

\bibitem{Abedi:2017jqx}
H.~Abedi and S.~Capozziello, ``{Gravitational waves in modified teleparallel
  theories of gravity},''
  \href{http://dx.doi.org/10.1140/epjc/s10052-018-5967-x}{{\em Eur. Phys. J.}
  {\bf C78} (2018) no.~6, 474},
\href{http://arxiv.org/abs/1712.05933}{{\tt arXiv:1712.05933 [gr-qc]}}.

\bibitem{Cai:2018rzd}
Y.-F. Cai, C.~Li, E.~N. Saridakis, and L.~Xue, ``{$f(T)$ gravity after GW170817
  and GRB170817A},'' \href{http://dx.doi.org/10.1103/PhysRevD.97.103513}{{\em
  Phys. Rev.} {\bf D97} (2018) no.~10, 103513},
\href{http://arxiv.org/abs/1801.05827}{{\tt arXiv:1801.05827 [gr-qc]}}.

\bibitem{Farrugia:2018gyz}
G.~Farrugia, J.~L. Said, V.~Gakis, and E.~N. Saridakis, ``{Gravitational Waves
  in Modified Teleparallel Theories},''
  \href{http://dx.doi.org/10.1103/PhysRevD.97.124064}{{\em Phys. Rev.} {\bf
  D97} (2018) no.~12, 124064},
\href{http://arxiv.org/abs/1804.07365}{{\tt arXiv:1804.07365 [gr-qc]}}.

\bibitem{Hohmann:2018xnb}
M.~Hohmann, ``{Polarization of gravitational waves in general teleparallel
  theories of gravity},''
  \href{http://dx.doi.org/10.1134/S1063772918120235}{{\em Astron. Rep.} {\bf
  62} (2018) no.~12, 890--897},
\href{http://arxiv.org/abs/1806.10429}{{\tt arXiv:1806.10429 [gr-qc]}}.

\bibitem{Hohmann:2018jso}
M.~Hohmann, M.~Krššák, C.~Pfeifer, and U.~Ualikhanova, ``{Propagation of
  gravitational waves in teleparallel gravity theories},''
  \href{http://dx.doi.org/10.1103/PhysRevD.98.124004}{{\em Phys. Rev.} {\bf
  D98} (2018) no.~12, 124004},
\href{http://arxiv.org/abs/1807.04580}{{\tt arXiv:1807.04580 [gr-qc]}}.

\bibitem{Conroy:2017yln}
A.~Conroy and T.~Koivisto, ``{The spectrum of symmetric teleparallel
  gravity},'' \href{http://dx.doi.org/10.1140/epjc/s10052-018-6410-z}{{\em Eur.
  Phys. J.} {\bf C78} (2018) no.~11, 923},
\href{http://arxiv.org/abs/1710.05708}{{\tt arXiv:1710.05708 [gr-qc]}}.

\bibitem{Runkla:2018xrv}
M.~Rünkla and O.~Vilson, ``{Family of scalar-nonmetricity theories of
  gravity},'' \href{http://dx.doi.org/10.1103/PhysRevD.98.084034}{{\em Phys.
  Rev.} {\bf D98} (2018) no.~8, 084034},
\href{http://arxiv.org/abs/1805.12197}{{\tt arXiv:1805.12197 [gr-qc]}}.

\bibitem{Adak:2005cd}
M.~Adak, M.~Kalay, and O.~Sert, ``{Lagrange formulation of the symmetric
  teleparallel gravity},''
  \href{http://dx.doi.org/10.1142/S0218271806008474}{{\em Int. J. Mod. Phys.}
  {\bf D15} (2006)  619--634},
\href{http://arxiv.org/abs/gr-qc/0505025}{{\tt arXiv:gr-qc/0505025 [gr-qc]}}.

\bibitem{Nester:1998mp}
J.~M. Nester and H.-J. Yo, ``{Symmetric teleparallel general relativity},''
  {\em Chin. J. Phys.} {\bf 37} (1999)  113,
\href{http://arxiv.org/abs/gr-qc/9809049}{{\tt arXiv:gr-qc/9809049 [gr-qc]}}.

\bibitem{Hoermander1}
L.~H{\"o}rmander, {\em The Analysis of Linear Partial Differential Operators
  {I}: Distribution Theory and {Fourier} Analysis}.
\newblock No.~256 in Grundlehren der mathematischen Wissenschaften. Springer,
  1983.

\bibitem{Hoermander2}
L.~H{\"o}rmander, {\em The Analysis of Linear Partial Differential Operators
  {II}: Differential Operators with Constant Coefficients}.
\newblock No.~257 in Grundlehren der mathematischen Wissenschaften. Springer,
  1983.

\bibitem{VanNieuwenhuizen:1973fi}
P.~Van~Nieuwenhuizen, ``{On ghost-free tensor lagrangians and linearized
  gravitation},''
\href{http://dx.doi.org/10.1016/0550-3213(73)90194-6}{{\em Nucl. Phys.} {\bf
  B60} (1973)  478--492}.

\bibitem{Newman:1961qr}
E.~Newman and R.~Penrose, ``{An Approach to gravitational radiation by a method
  of spin coefficients},''
\href{http://dx.doi.org/10.1063/1.1724257}{{\em J. Math. Phys.} {\bf 3} (1962)
  566--578}.

\bibitem{Will:1993ns}
C.~M. Will, {\em {Theory and experiment in gravitational physics}}.
\newblock
1993.
\newblock

\end{thebibliography}\endgroup
\end{document}